# Micius, the world's first quantum communication satellite, was hackable

Alexander Miller
*Department of Physics*
*National University of Singapore*
Singapore, Singapore
ORCID 0000-0003-2174-4431

*Abstract*—The decoy-state BB84 protocol for quantum key distribution (QKD) is used on Micius, the world's first satellite for quantum communications. The method of decoy states can detect photon-number-splitting eavesdropping and thus enables, in theory, secure QKD using weak coherent pulses over long distances with high channel loss inherent in satellite communication systems. However, it is widely known that realistic QKD devices can be vulnerable to various types of side-channel attacks that rely on flaws in experimental implementation. In most free-space QKD systems, including that on board Micius, multiple semiconductor lasers with passive optics are utilized to randomly generate polarization states. Optical pulses from independent laser diodes can to some extent differ in their temporal, spectral, and/or spatial distribution, and the quantum states can thus be distinguishable. Such distinguishability of photons in additional non-operational degrees of freedom compromises unconditional security of QKD since an eavesdropper can exploit this loophole to improve their attack strategy. The author carried out a thorough analysis of the experimental data obtained during multiple communication sessions between Micius and one of the ground stations designed specifically for it. Relative time delays between all the eight laser diodes on board have been found. The typical desynchronization between the lasers exceeds 100 ps, which is comparable with their pulse duration of 200 ps. The largest time delay was observed between signal and decoy states for vertically polarized photons and is about 300 ps. With such mismatch in timing, a potential attacker using as perfect equipment as possible unless it violates the laws of physics was shown to be capable of distinguishing decoy states from signal ones in at least 98.7% of cases. It offers great potential for hacking since one of the assumptions that the security of the decoy-state protocol is based on is that intensity levels selected for transmission of qubits are not known in advance. Based on a prior theoretical study, where a photon number splitting attack strategy that leverages the benefits of distinguishable decoy states is proposed, the author shows that distribution of quantum keys from Micius was insecure.

I. INTRODUCTION

The idea of quantum cryptography is not based on computational complexity of mathematical algorithms, but on the laws of physics. The idea was first announced in 1984 by Bennett and Brassard, who proposed the first quantum cryptography protocol, BB84 [1]. An experimental implementation of quantum key distribution was first demonstrated in 1989 [2]. In that experiment, qubits were encoded in photon's polarization and the photons were transmitted through 32.5 cm on an optical table. Since then, the technology of quantum communications has been widely developed [3]. The first security proofs of QKD protocols, such as BB84, imply that the sender, Alice, uses a perfect source that only emits single photons [4,5]. However, real QKD systems utilize practical sources, such as weak coherent state laser source, which can generate multi-photon states with non-zero probability. An eavesdropper, Eve, can split such multi-photon components, keep one photon and send the rest to the recipient, Bob. Once Alice and Bob disclose information about the bases, Eve can measure the states of intercepted photons and gain some information of the key. In practice, most quantum channels are lossy, and Eve can implement more sophisticated photon number splitting (PNS) attacks [6]. In order to ensure security when distributing quantum keys Alice has to use an extremely weak laser source, which obviously leads to significant reduction in key rate or giving up the possibility of QKD over large distances [7,8]. In 2002, Hwang proposed the method of decoy states, which allows to address the issue of security while using multi-photon sources [9]. The essence of the method is to use different randomly chosen intensity levels, one for signal states and the others for so called decoy states. After Bob measures the states of photon qubits, Alice announces which intensity level has been chosen for the transmission of each one. By monitoring quantum bit error rate (QBER) associated with each intensity level, the two legitimate parties will be able to detect a PNS attack. Lo, Ma and their coauthors, and Wang, developed the idea of Hwang, gave a rigorous security proof and suggested protocols with finite numbers of decoy states [10–12]. They were one-decoy-state protocol, which was used in the first experimental implementation [13], and two-decoy-state protocol. Later, two-decoy-state protocol became the most widely used in real QKD devices due to the optimal combination of its relative ease of implementation and high secret key rate. Its particular case, vacuum+weak decoy-state BB84 protocol is used on Micius, the world's first quantum communication satellite [14].

Although the method of decoy states prevents the threat of PNS attack, there can be non-operational degrees of freedom, which are not considered in theory, making the states distinguishable. Attacks that take advantage of these additional degrees of freedom are called side-channel attacks [15–22]. Loopholes on receiver's side can be closed with the use of measurement-device-independent QKD protocol [23], where measurement devices are excluded from the private space of legitimate sides and transferred to a third party. However, this protocol does not close loopholes on transmitter's side. A malicious party can extract some information of the key using light source side channels. In particular, photons in different states can be distinguished by their emission time [24,25] and/or spectrum [24]. In free-space QKD, there can also be spatial distinguishability due to angular misalignment between the laser sources [24,26]. The transmitter source at Micius consists of eight separate laser diodes: four types of polarization for signal states and four ones for decoy states. There can thus be distinguishability in all the above-mentioned degrees of freedom. Previously, the lasers on board Micius were reported to be synchronized with accuracy, not worse than 10 ps, which is much smaller than their pulse duration of 200 ps [14]. The author conducted a thorough independent analysis of the experimental data obtained during multiple communication sessions between Micius and one of the optical ground stations developed specifically for it. The experiments were carried out from October 2021 to March 2022. Relative time delays between all the laser diodes on board as they were during those experiments have been found. The author concludes whether distribution of quantum keys from Micius had been really secure. Finally, they provide some recommendations on how such QKD systems could be improved in the future.

## II. Experiment

### A. Micius

The world's first quantum communication satellite, Micius, was launched into its orbit in 2016, which was followed by experiments on satellite-to-ground QKD [14,27,28], entanglement distribution [29], entanglement-based QKD [30], and quantum teleportation [31]. The optical design of its payload for prepare-and-measure QKD is presented in Fig. 1.

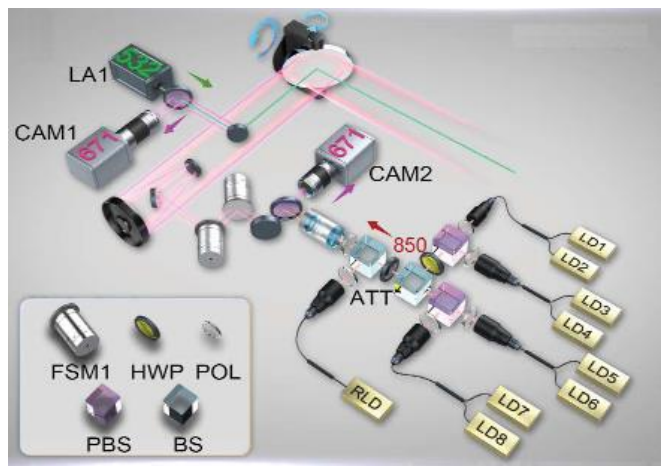

Fig. 1. Optical design of the quantum transmitter at Micius [14]: LD1–LD8 – 850 nm laser diodes, POL – polarizing filters, PBS – polarization beam splitters, HWP – half-wave plates, BS – beam splitters, ATT – fixed optical attenuators, LA1 – beacon laser, CAM1 – WFOV camera, CAM 2 – NFOV camera, FSM – fast steering mirrors, RLD – polarization reference laser diode.

Laser pulses generated by eight separated laser diodes LD1–LD8 pass through a standard BB84 encoding module that consists of two polarization beam splitters PBS, a half-wave plate HWP, and a 50:50 beam splitter BS, and are attenuated by an optical attenuator ATT. The combined signal is co-aligned with a green laser (LA1) beam for the tracking system and time synchronization and is sent out through a 300 mm aperture Cassegrain telescope. The payload is also equipped with one more 850 nm laser RLD, whose power is not attenuated to a single photon level and which is used as a polarization reference. A two-axis gimbal mirror in the output of the telescope and a wide field-of-view (WFOV) camera CAM1 are combined for coarse tracking loop control. Two fast steering mirrors FSM and a narrow filed-of-view (NFOV) camera are used for fine tracking. Independent precise electrical control of the eight lasers allows to obtain the required average photon numbers at the output of the telescope: 0.8 and 0.1 for signal and for weak states, respectively. Vacuum states correspond to the absence of triggering of any of the eight laser diodes. Signal, weak, and vacuum states are sent with probabilities of 50%, 25%, and 25%, correspondingly. The intensity level as well as polarization chosen to prepare a photon state is determined by four random bits, whose sequence is generated by a physical thermal noise device. More details about Micius can be found in [14].

### B. Optical Ground Station

The optical ground station (OGS) is placed at Zvenigorod observatory in Russia (55°41′56″N, 36°45′32″E, 180 m above msl) and based on a 60 cm aperture telescope equipped with a motorized tracking mount. Its scheme is presented in Fig. 2.

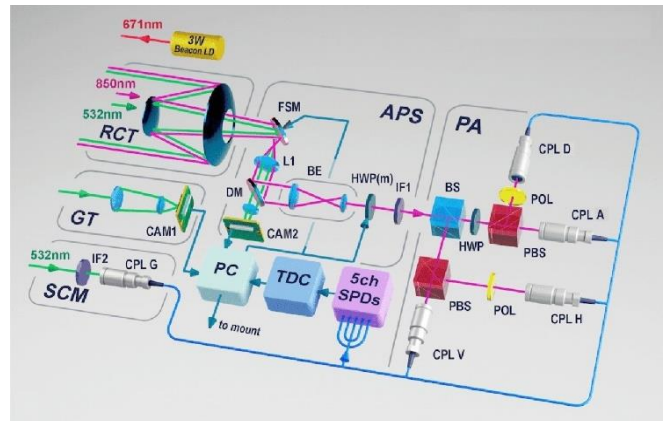

Fig. 2. Schematic of the optical ground station: RCT – Ritchey-Chretien telescope, GT – guide telescope, CAM1 – WFOV camera, SCM – synchronization module, IF – interference filter, CPL – fiber optic collimators, APS – analysis and processing system, FSM – fast steering mirror, L1 – focusing lens, DM – dichroic mirror, CAM2 – NFOV camera, BE – beam expander, HWP – half-wave plates, PC – personal computer, TDC – time-to-digital converter, SPD – single photon detectors, PA – polarization analyzer, BS – 50:50 beam splitter, PBS – polarization beam splitters, POL – polarizing filters.

The primary purpose of the telescope system is collection of the quantum signal. It is capable of determining the trajectory of a satellite and tracking it regardless of whether a downlink beacon signal is captured or not. Knowledge of the satellite's trajectory is obtained by means of SGP4 model, which converts a two-line element set of the satellite to its celestial coordinates fed to the telescope mount. In practice, the typical RMS-error does not exceed 400 µrad when tracking low-orbit satellites. A further reduction in pointing error is achieved through the use of an acquisition, pointing, and tracking system, which consists of coarse and fine tracking system, and an uplink beacon. The uplink beacon (671 nm) helps Micius to orientate itself at the OGS and establish a good quality line-of-sight. The ground-based station is equipped with a WFOV camera CAM1, whose sensor is in the focal plane of a guide telescope GT mounted parallel to the optical axis of the main one. On-the-fly processing of image stream from the WFOV camera is permanently carried out and, once the camera reliably receives the downlink beacon signal (532 nm), a required offset starts to be calculated. The offset is then supplied to the mount control software, which allows to minimize pointing error yet remaining due to imperfect knowledge of the satellite's orbit and compensate any systematic tracking errors inherent in the telescope mount. A typical pointing error at closed-loop coarse tracking is of about 50 µrad RMS. The OGS is also equipped with a NFOV camera CAM2 and a fast steering mirror FSM. Unlike the WFOV one, the NFOV camera and the FSM are mounted inside the optical receiver. The FSM is introduced into the combined beam path of the quantum and downlink beacon signal to guide the quantum signal into detector's field of view. After locating Micius in the NFOV camera, the remaining pointing error is compensated by beam steering. A typical RMS-error during closed-loop fine tracking does not exceed 7 μrad. The quantum part of the receiver is a standard BB84 decoding module with passive basis selection that consists of a 50:50 beam splitter BS, a half-wave plate HWP, and two polarization beam splitters PBS. The quantum signal is coupled into 105 μm core multi-mode optical fiber and then falls on single photon detectors SPDs.

*C. Method of Determining Laser Diode Time Delays*

The method of determining time delays between laser diodes is based on comparison of the response times from the corresponding quantum states at a chosen detector. Since the detector is the same, the factor of different optical path lengths is completely eliminated and discrepancy in response time can only be caused by a mismatch in firing time of lasers. Let $P$ denote one of four polarization states, horizontal ($H$), vertical ($V$), linear polarization at an angle of +45° ($D$) and -45° ($A$), and subscript μ denote one of the two intensity levels at the transmitter's sourse, signal ($s$) and decoy ($d$). Let $i$ and $j$ denote sequence numbers of two arbitrary quantum states $X$ and $Y$, which belong to set $\{P_\mu\}$. Then their times of detection, $t_{d,i}^X$ and $t_{d,j}^Y$, can be written as follows:

$$t_{d,i}^X = t_{e,i}^X + t_p + \delta t_i^{rand} \quad (1)$$

and

$$t_{d,j}^Y = t_{e,i}^Y + t_p + \delta t_j^{rand}, \quad (2)$$

where $t_e$ is emission time at the transmitter, $t_p$ is signal propagation time, and $\delta t^{rand}$ is a random deviation, which is caused by timing jitter of the detectors, atmospheric scintillation and some other factors. Signal propagation time $t_p$ is assumed to be independent of which quantum state is sent. So is the random error $\delta t^{rand}$. Both these values can only depend on the detector that registers the signal.

Emission times can be represented as

$$t_{e,i}^X = \delta t_e^X + T \cdot i \quad (3)$$

and

$$t_{e,i}^Y = \delta t_e^Y + T \cdot j, \quad (4)$$

where $\delta t_e^X$ and $\delta t_e^Y$ are some constant time delays of states $X$ and $Y$, and $T$ is the repetition period of qubits.

By subtracting (1) from (2), if taking into account (3) and (4), one can obtain:

$$t_{d,j}^Y - t_{d,i}^X + T \cdot (i - j) = \delta t_e^Y - \delta t_e^X + \delta t_j^{rand} - \delta t_i^{rand}. \quad (5)$$

$\delta t_i^{rand}$ and $\delta t_j^{rand}$ are random variables, whose mathematical expectations can reasonably be made equal to zero. Consequently, the expression on the left side of (5) is also a random variable, and its expected value in only determined by $\delta t_e^Y - \delta t_e^X$. In accordance with the properties of mathematical expectation, the following equation can be obtained from (5):

$$E(t_{d,j}^Y - T \cdot j) - E(t_{d,i}^X - T \cdot i) = \delta t_e^Y - \delta t_e^X \quad (6)$$

Let $\delta t(Y, X)$ denote the difference between $\delta t_e^Y$ and $\delta t_e^X$ for short. The physical meaning of this value is the degree of desynchronization between lasers $X$ and $Y$. The mathematical expectations of $t_{d,i}^X - T \cdot i$ and $t_{d,j}^Y - T \cdot j$, unlike the values themselves, no longer depend on indices $i$ and $j$. The expected values can be denoted as $t(X \rightarrow Z)$ and $t(Y \rightarrow Z)$ correspondingly, where $Z$ denotes any chosen detector. Then the value of desynchronization can be written as follows:

$$\delta t(Y, X) = t(Y \rightarrow Z) - t(X \rightarrow Z). \quad (7)$$

For example, the mismatch in firing time between lasers $V_d$ and $V_s$ should be sought by analyzing the count statistics of detector $V$ since it obviously gets the most photons having vertical polarization, which ensures the best signal-to-noise ratio in this case:

$$\delta t(V_d, V_s) = t(V_d \rightarrow V) - t(V_s \rightarrow V) \quad (8)$$

The corresponding distributions of detection time obtained on October 31, 2021 are presented in Fig. 3.

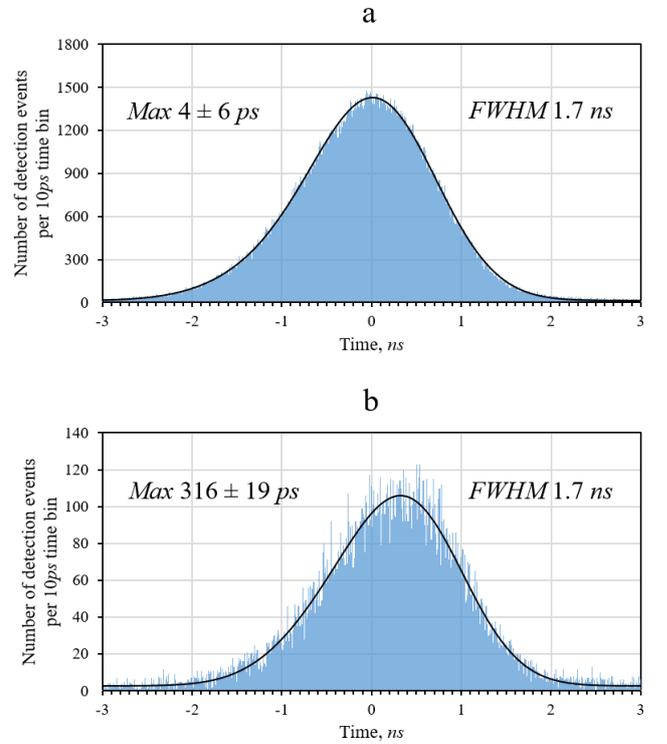

Fig. 3. Distribution of detection time of detector V: Alice sent $V_s$ (a) and $V_d$ (b). Detection events in the time interval from UTC2021-10-31T22.37.53 to UTC2021-10-31T22.39.30 (97 s) are analyzed.

The histograms show the number of detection events per 10 ps time bin, whereas the black curves represent the results of approximation by the least squares method. The maximum point of probability density falls on 4 and 316 ps for signal and decoy states correspondingly. The value of 3σ, where σ is the square root of the variance, is of 6 ps for $V_s$ and 19 ps for $V_d$. The variance, $\sigma^2$, is mainly determined by signal-to-noise ratio – the details on how the error of approximation is calculated are given in Appendix A. Since the obtained distributions correspond to independent random variables, when subtracting $t(V_s \rightarrow V)$ from $t(V_d \rightarrow V)$, their variances are summed. It is easy to calculate that 3σ for $\delta t(V_d, V_s)$ is of about 20 ps. Thus, the mismatch in firing time between laser diodes $V_d$ and $V_s$ is of 312 ± 20 ps. In accordance with Bienayme-Chebyshev inequality, the probability that a random variable deviates from its mean by more than 3σ is at most ⅑. Analysis of the actual experimental data shows that the probability that the true value is outside 3σ interval is of about 0.023.

The other time delays are sought similarly to $\delta t(V_d, V_s)$. In this study, relative delays between the laser diodes are determined. Consequently, without loss of generality the time delay of one of the lasers can be assumed to be equal to zero. In this work, $H_s$ was chosen as the laser with zero delay, whereas the other delays are calculated relative to it. All the equations used to calculate time delays and their derivation are given in Appendix B. An overview of the main results obtained in the experiments with Micius is presented in the next section.

### III. RESULTS

The time delays between laser diodes on board Micius as they were on October 31, 2021, March 9, 2022, and March 23, 2022, are given in Table 1 and also presented in Fig. 4. The typical degree of desynchronization defined as the unbiased estimate of variance with unknown mean exceeds 100 ps. The maximum mismatch in time was observed between signal and decoy states for vertically polarized photons and is about 300 ps. As one can notice, all the time delays are highly stable: they stay pretty much the same throughout the months. As can be seen, the widths of the error bars, which represent 3σ intervals, significantly change from experiment to experiment. The magnitude of the error is mainly determined by signal-to-noise ratio, which in turn depends primarily on two factors: the satellite elevation angles during a communication session and the time it lasts. Besides that, the condition of the atmosphere also impacts total channel loss and thus signal-to-noise ratio. The duration of the QKD session carried out on October 31, 2021, March 9, 2022, and March 23, 2022, is 170, 261, and 220 s, correspondingly. However, only detection events from shortened time intervals of 97 and 202 s are taken into consideration for the data obtained on October 31, 2021 and March 23, 2022, respectively, due to technical difficulties encountered during processing.

### IV. ATTACK STRATEGY

It was previously reported that the width of optical pulses at Micius is 200 ps [14]. In this work, the intensity distribution over time is assumed to be normal with FWHM of 200 ps. The corresponding intensity over time for optical pulses $V_s$ and $V_d$, with a delay between them of 312 ps, as it was on October 31, 2021, is presented in Fig. 5.

A potential violator can apply the following strategy to attack the QKD system under discussion. They assign two time gates: one to signal states and the other one to decoy states. Then they intercept photons sent from Alice to Bob and infer about which state have been caught from trigger time. The events that are covered by neither of the gates can merely be discarded. The minimum gate width that can be set by the eavesdropper is chosen due to the assumption that they have to receive equal or greater share of useful signal than Bob does. During the experiments on QKD with Micius, spatial, spectral, and temporal filtering was applied to reduce noise primarily resulted from background radiation collected by the telescope along with quantum signal.

TABLE I. LASER DIODE TIME DELAYS AT MICIUS.

| Laser Diode | Time Delay Relative to Laser Diode $H_s$, ps | | |
|---|---|---|---|
| | *October 31, 2021* | *March 9, 2022* | *March 23, 2022* |
| $V_s$ | −10 ± 14 | 12 ± 8 | −4 ± 10 |
| $D_s$ | 29 ± 50 | 33 ± 29 | 5 ± 49 |
| $A_s$ | 139 ± 54 | 157 ± 23 | 168 ± 35 |
| $H_d$ | 246 ± 25 | 224 ± 13 | 220 ± 16 |
| $V_d$ | 302 ± 23 | 320 ± 12 | 293 ± 17 |
| $D_d$ | 223 ± 42 | 197 ± 24 | 168 ± 39 |
| $A_d$ | 176 ± 50 | 173 ± 20 | 190 ± 31 |

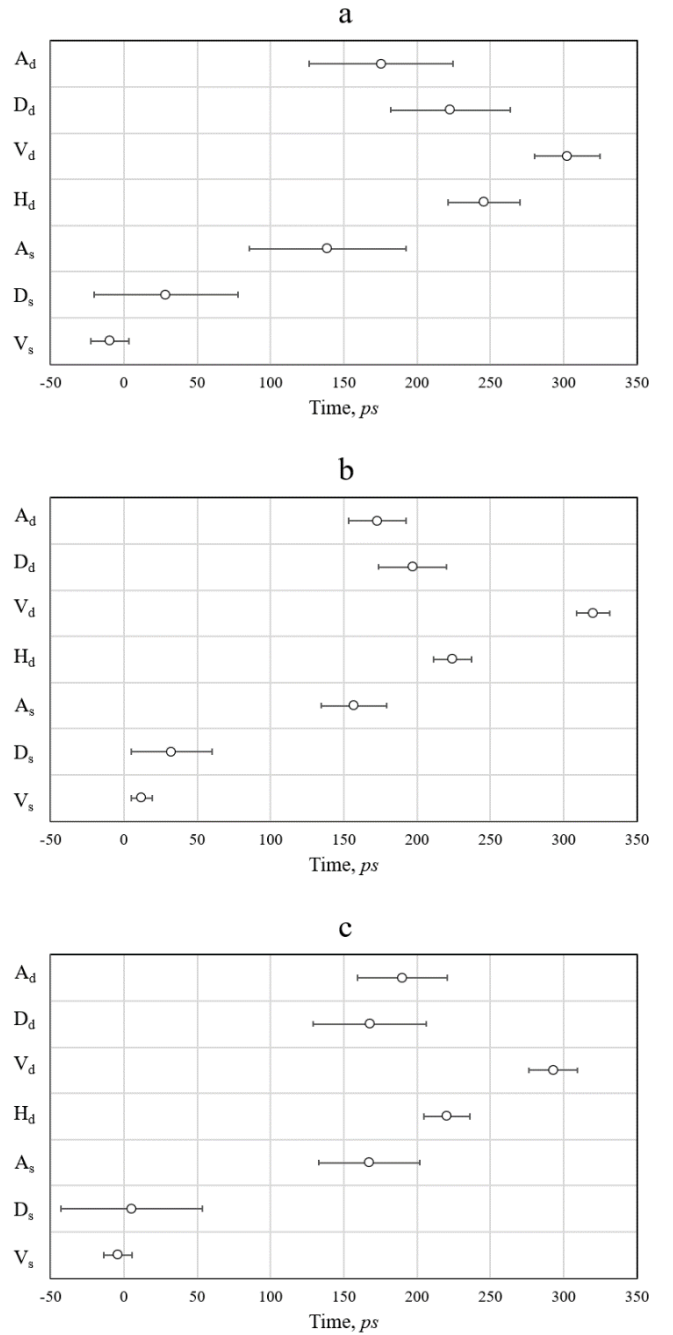

Fig. 4. Laser diode time delays relative to $H_s$ at Micius on October 31, 2021 (a), March 9, 2022 (b), and March 23, 2022 (c). The error bars represent ±3σ intervals, which cover the true values with a probability of 0.977.

In the experiments under consideration, the time gate width was set to 2 ns, which allowed to maintain QBER at sufficiently low level, less than 2%, during a whole communication session. On October 31, 2021, synchronization precision was about 720 ps (σ), or 1.7 ns FWHM. If assuming normal distribution, it can be calculated that Bob received about 83.5% of useful signal at a gate width of 2 ns. The eavesdropper is supposed to possess as advanced equipment as possible unless it is forbidden by the laws of physics. In particular, they have detectors with as low jitter as possible up to zero. At zero jitter, the eavesdropper may set the gate width to 235 ps and obtain the same 83.5% of useful signal. Based on detection time, they make a suggestion on which intensity level had been chosen by Alice to prepare an intercepted photon. The probability that the violator fails in their guess is determined by area under the gaussian tails outside the gates, which is denoted as red color areas in Fig. 5. By integration of the red areas, it can be calculated that the probability of failure is 1.3%. In other words, it means that the eavesdropper is able to correctly determine the intensity level selected for transmission of a qubit in 98.7% of cases. It provides ample opportunities for hacking since the absolute indistinguishability of quantum states in any non-operational degree of freedom is one of the assumptions that the security of a QKD protocol is based on.

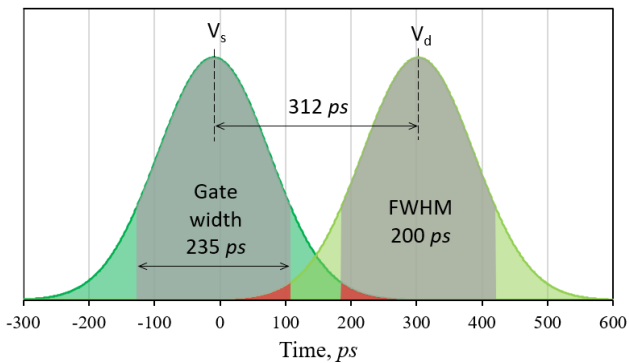

Fig. 5. Intensity over time for optical pulses $V_s$ and $V_d$ at Micius, and a possible attack scheme of an intruder.

It should be noted that the probability of the intruder making a mistake of 1.3% should be considered as rather an upper bound. In practice, any quantum communication channel is lossy. Also, a potential attacker may use detectors with a quantum efficiency higher than that of Bob's detectors. It would allow them to narrow the time gates even more and thus reduce the probability of failure. However, determining the optimal gate width is beyond the scope of this work and can be a subject of a separate study.

Previously, Huang *et al.* proposed an explicit scheme of a PNS attack that takes advantage of distinguishability of decoy states [25]. Using their theoretical model, the author made estimate calculations based on the real parameters of the experiments with Micius. The secret key rate was shown to be zero at a temporal mismatch between signal and decoy states of 300 ps.

## V. Conclusions

The author has carried out a thorough analysis of the experimental data obtained during multiple communication sessions with Micius, the world's first quantum communication satellite, and found relative time delays between all the laser diodes within the on-board quantum transmitter. It was found that the typical desynchronization between the lasers exceeds 100 ps, which is comparable with their pulse duration of 200 ps. The largest time delay was observed between signal and decoy states for vertically polarized photons and is about 300 ps. At such mismatch in time, a potential eavesdropper was shown to be capable of distinguishing decoy states from signal ones in at least 98.7% of cases. It opens great opportunities for hacking since one of the assumptions that the security of the decoy-state protocol is based on is that intensity levels selected by Alice for transmission of qubits are not known in advance. Based on a prior theoretical study [25], where an explicit scheme of PNS attack that takes advantage of distinguishable decoy states is proposed, the author carried out a security analysis of the QKD system under consideration at the real experimental parameters. At a temporal mismatch in time between signal and decoy states of 300 ps, the secret key rate was shown to be zero. Thus, distribution of quantum keys from Micius was insecure.

It should be noted that only distinguishability between signal and decoy states is considered in [25], whereas distinguishability among different BB84 states is not taken into account. A more general theoretical model that takes into consideration an arbitrary distribution pattern is required to enable a more rigorous security analysis. To the author's knowledge, there has not been such a holistic study so far. It should also be mentioned that this work considers neither spectral nor spatial distinguishability of photons, which could become topics of future studies.

The laser diodes on board Micius were previously reported to be synchronized with an accuracy of 10 ps [14]. The question why the new data contradicts that statement has remained open. Hypothetically, something wrong could have happened to the on-board quantum transmitter by October 2021. However, as reported in [14], all the lasers are controlled independently and it seems therefore highly unlikely that a malfunction had happened to all the drivers at once. Moreover, as observed in this study, all the time delays are highly stable: within the margin of error, they stay the same for several months. This finding suggests that the delays in earlier experiments [14,27,28] could have been the same or very close to those observed from October 2021 to March 2022. Anyway, a command to correct the time delays from Earth had not apparently been envisaged in this mission.

This work once again shows how insecure realistic devices for quantum communication can be despite theoretically proven security of the used QKD protocols. When exploiting quantum transmitters based on multiple lasers, one should pay due attention to proper time synchronization and the possibility to operatively configure time delays. It is of special importance in space missions, where changing parameters into orbit can be somewhat complicated and the price of an error can be extremely high. Rigorous pre-flights tests should be performed by the launch of a spacecraft. Besides that, special commands to control temperature, current, time delays, and other parameters, from Earth should be envisaged. The use of a QKD transmitter based on a single laser and electro-optical modulators can be an alternative to standard multiple laser-based source. Nevertheless, using a single laser-based transmitter does not guarantee the complete elimination of side channels. In particular, pump-current modulation can result in a side channel in time domain as well, which allows an eavesdropper to distinguish decoy states from signal ones [25]. Quantum sources based on the use of an electro-optical intensity modulator can also be leaky [17]. It should also be noted that using electro-optical modulators makes a QKD system more difficult to implement from a technical point of view, especially at high rates. Entanglement-based QKD [29,30,32,33] is one more of possible solutions. It enables secure communications in trusted node-free networks and is not vulnerable to PNS attacks. However, the presence of side channels in entanglement-based QKD systems cannot be fully excluded too.

ACKNOWLEDGMENT

The author is grateful to QSpace Technologies for providing data obtained during their experiments with Micius. The author also thanks Vadim Makarov for fruitful discussions.

REFERENCES

[1] C. H. Bennett and G. Brassard, "Quantum cryptography: public key distribution and coin tossing", Proc. Int. Conf. of Computers, Systems & Signal Processing, Bangalore, India, Dec. 9–12, 1984, vol. 1, pp. 175–179.
[2] C. H. Bennett, F. Bessette, G. Brassard, L. Salvail, and J. Smolin, "Experimental quantum cryptography", J. Cryptol., vol. 5, pp. 3–28, 1992.
[3] S. Pirandola *et al.*, "Advances in quantum cryptography", Adv. Opt. Photon., vol. 12, pp. 1012–1236, 2020.
[4] D. Mayers, "Quantum key distribution and string oblivious transfer in noisy channels", Adv. Cryptol. – CRYPTO'96, pp. 343–357, 1996.
[5] P. W. Shor and J. Preskill, "Simple proof of security of the BB84 quantum key distribution protocol", Phys. Rev. Lett., vol. 85, p. 441, 2000.
[6] G. Brassard, N. Lutkenhaus, T. Mor, and B. C. Sanders, "Limitations on practical quantum cryptography", Phys. Rev. Lett., vol. 85, p. 1330, 2000.
[7] D. Gottesman, H.-K. Lo, N. Lutkenhaus, and J. Preskill, "Security of quantum key distribution with imperfect devices", Quantum Inf. Comput., vol. 4, pp. 325–360, 2004.
[8] H.-K. Lo and J. Preskill, "Security of quantum key distribution using weak coherent states with nonrandom phases", Quantum Inf. Comput., vol. 7, pp. 431–458, 2007.
[9] W.-Y. Hwang, "Quantum key distribution with high loss: toward global secure communication", Phys. Rev. Lett., vol. 91, p. 057901, 2003.
[10] H.-K. Lo, X. Ma, and K. Chen, "Decoy state quantum key distribution", Phys. Rev. Lett., vol. 94, p. 230504, 2005.
[11] X. Ma, B. Qi, Y. Zhao, and H.-K. Lo, "Decoy state quantum key distribution", Phys. Rev. A, vol. 72, p. 012326, 2005.
[12] X.-B. Wang, "Beating the photon-number-splitting attack in practical quantum cryptography", Phys. Rev. Lett., vol. 94, p. 230503, 2005.
[13] Y. Zhao, B. Qi, X. Ma, H.-K. Lo, and L. Qian, "Experimental quantum key distribution with decoy states", Phys. Rev. Lett., vol. 96, p. 070502, 2006.
[14] S.-K. Liao *et al.*, "Satellite-to-ground quantum key distribution", Nature, vol. 549, pp. 43–47, 2017.
[15] V. Scarani and C. Kurtsiefer, "The black paper of quantum cryptography: real implementation problems", Theor. Comput. Sci., vol. 560, pp. 27–32, 2014.
[16] N. Jain, B. Stiller, I. Khan, D. Elser, C. Marquardt, and G. Leuchs, "Attacks on practical quantum key distribution systems (and how to prevent them)", Contemp. Phys., vol. 57, pp. 366–387, 2016.
[17] K. Tamaki, M. Curty, and M. Lucamarini, "Decoy-state quantum key distribution with a leaky source", New J. Phys., vol. 18, p. 065008, 2016.
[18] M. Pereira, M. Curty, and K. Tamaki, "Quantum key distribution with flawed and leaky sources", Npj Quantum Inf., vol. 5, p. 62, 2019.
[19] F. Xu, X. Ma, Q. Zhang, H.-K. Lo, and J.-W. Pan, "Secure quantum key distribution with realistic devices", Rev. Mod. Phys., vol. 92, p. 025002, 2020.
[20] A. Duplinskiy and D. Sych, "Bounding passive light-source side channels in quantum key distribution via Hong-Ou-Mandel interference", Phys. Rev. A, vol. 104, p. 012601, 2021.
[21] D. Babukhin, D. Kronberg, and D. Sych, "Explicit attacks on the Bennett-Brassard 1984 protocol with partially distinguishable photons", Phys. Rev. A, vol. 106, p. 042403, 2022.
[22] P. Chaiwongkhot *et al.*, "Eavesdropper's ability to attack a free-space quantum-key-distribution receiver in atmospheric turbulence", Phys. Rev. A, vol. 99, p. 062315, 2019.
[23] H.-K. Lo, M. Curty, and B. Qi, "Measurement-device-independent quantum key distribution", Phys. Rev. Lett., vol. 108, p. 130503, 2012.
[24] S. Nauerth, M. Furst, T. Schmitt-Manderbach, H. Weier, and H. Weinfurter, "Information leakage via side channels in freespace BB84 quantum cryptography", New J. Phys., vol. 11, p. 065001, 2009.
[25] A. Huang, S.-H. Sun, Z. Liu, and V. Makarov, "Quantum key distribution with distinguishable decoy states", Phys. Rev. A, vol. 98, p. 012330, 2018.
[26] P. Arteaga-Diaz, D. Cano, and V. Fernandez, "Practical side-channel attack on free-space QKD systems with misaligned sources and countermeasures", IEEE Access, vol. 10, pp. 82697–82705, 2022.
[27] S.-K. Liao *et al.*, "Satellite-relayed intercontinental quantum network", Phys. Rev. Lett., vol. 120, p. 030501, 2018.
[28] Y.-A. Chen *et al.*, "An integrated space-to-ground quantum communication network over 4,600 kilometres", Nature, vol. 589, pp. 214–219, 2021.
[29] J. Yin *et al.*, "Satellite-based entanglement distribution over 1200 kilometers", Science, vol. 356, pp. 1140–1144, 2017.
[30] J. Yin *et al.*, "Entanglement-based secure quantum cryptography over 1120 kilometers", Nature, vol. 582, pp. 501–505, 2020.
[31] J.-G. Ren *et al.*, "Ground-to-satellite quantum teleportation", Nature, vol. 549, pp. 70–73, 2017.
[32] Z. Tang *et al.*, "Generation and analysis of correlated pairs of photons aboard a nanosatellite", Phys. Rev. Applied, vol. 5, p. 054022, 2016.
[33] A. Villar *et al.*, "Entanglement demonstration on board a nano-satellite", Optica, vol. 7, pp. 734–737, 2020.

APPENDIX A

ERROR IN DETERMINING THE MAXIMUM OF PROBABILITY DENSITY FUNCTION WHEN APPROXIMATING IT FROM NOISY DATA BY THE LEAST SQUARES METHOD

Let $f$ be a function of a single variable, $x$, that has a set of parameters, $\{P\}$: $f = f(x,\{P\})$. Let the measurement results of quantity $f$ at some finite set of points, $\{x_i\}$, $i = 1 \ldots n$, be known. As there are always some measurement errors, the obtained values, $f_i$, differ from the true ones: $f_i = f(x_i,\{P\}) + \delta f_i$, where $\delta f_i$ are the measurement errors. Using the least squares method, such parameters $\{P_a\}$ can be found so that the deviations of $f(x_i,\{P_a\})$ from $f_i$ are minimal in the aggregate. The question arises of to what extend the parameters thus found differ from the actual parameters, $\{P\}$. Or, specifying the issue, if $\delta f_i$ are random variables, what the variance of each of the values from $\{P_a\}$ is. This work answers the question for the special case of one-dimensional Gaussian function, which is defined by three parameters – $A$, $\mu$, and $\sigma$: $f(x) = A \cdot e^{-\frac{(x-\mu)^2}{2\sigma^2}}$.

Since the least squares method is used for approximation, the local extremum of function $\sum_{i=1}^{n}\bigl(f(x_i,\{P\}) + \delta f_i - f(x_i,\{P_a\})\bigr)^2$ is sought. In the case of a Gaussian function, it can thus be written:

$$\frac{\partial}{\partial p_j}\sum_{i=1}^{n}\left(A \cdot e^{-\frac{(x_i-\mu)^2}{2\sigma^2}} + \delta f_i - A_a \cdot e^{-\frac{(x_i-\mu_a)^2}{2\sigma_a^2}}\right)^2 = 0 \qquad (1)$$

where $p_j$ are parameters from set $\{A, \mu, \sigma\}$.

In the context of the present study, only the difference between $\mu_a$ and $\mu$ is of interest. It can therefore be assumed that only $\mu_a$ varies, whereas the other two parameters are known: $A_a = A$ and $\sigma_a = \sigma$. Then (1) is transformed into:

$$\frac{d}{d\mu_a}\sum_{i=1}^{n}\left(A \cdot e^{-\frac{(x_i-\mu)^2}{2\sigma^2}} + \delta f_i - A \cdot e^{-\frac{(x_i-\mu_a)^2}{2\sigma^2}}\right)^2 = 0 \quad (2)$$

Taking the derivative and multiplying the result by $-\frac{\sigma}{2A}$, one can get the following equation:

$$\sum_{i=1}^{n}\frac{x_i-\mu_a}{\sigma}\cdot e^{-\frac{(x_i-\mu_a)^2}{2\sigma^2}} \cdot$$

$$\cdot\left(A \cdot e^{-\frac{(x_i-\mu)^2}{2\sigma^2}} + \delta f_i - A \cdot e^{-\frac{(x_i-\mu_a)^2}{2\sigma^2}}\right) = 0 \quad (3)$$

If we introduce notations $z_i = \frac{x_i-\mu}{\sigma}$ and $\varepsilon = \frac{\mu_a-\mu}{\sigma}$, (3) can be written as:

$$\sum_{i=1}^{n}(z_i - \varepsilon)\cdot e^{-\frac{1}{2}(z_i-\varepsilon)^2} \cdot$$

$$\cdot\left(A \cdot e^{-\frac{1}{2}z_i^2} + \delta f_i - A \cdot e^{-\frac{1}{2}(z_i-\varepsilon)^2}\right) = 0 \quad (4)$$

One can consider the case of small variations in $\mu$: $\varepsilon \ll 1$. In this work, perturbations $\delta f_i$ are sufficiently small and this condition is satisfied. By expanding the sum in (4) into Taylor series about $\varepsilon$ and only keeping linear terms, one may obtain:

$$\sum_{i=1}^{n}z_i\cdot e^{-\frac{1}{2}z_i^2}\cdot \delta f_i =$$

$$= \varepsilon\cdot\sum_{i=1}^{n}\left(A\cdot z_i^2\cdot e^{-z_i^2} + (1-z_i^2)\cdot e^{-\frac{1}{2}z_i^2}\cdot \delta f_i\right) \quad (5)$$

As small perturbations ($|\delta f_i| \ll A\cdot e^{-\frac{1}{2}z_i^2}$) are considered, the terms with $\delta f_i$ on the right side of (4) can be neglected and the following expression can be obtained for $\varepsilon$:

$$\varepsilon = \frac{\sum_{i=1}^{n}z_i\cdot e^{-\frac{1}{2}z_i^2}\cdot \delta f_i}{A\cdot \sum_{i=1}^{n}z_i^2\cdot e^{-z_i^2}} \quad (6)$$

The question arises of what the variance of $\varepsilon$ as a random variable is. Using well-known properties of variance and making the assumption that deviations $\delta f_i$ are independent, uncorrelated random values, one can get:

$$D(\varepsilon) = \frac{\sum_{i=1}^{n}z_i^2\cdot e^{-z_i^2}D(\delta f_i)}{A^2\cdot\left(\sum_{i=1}^{n}z_i^2\cdot e^{-z_i^2}\right)^2} \quad (7)$$

One can also suppose that random deviations $\delta f_i$ are of the same nature for all the points within the entire measuring range. Then $D(\delta f_1) = \cdots = D(\delta f_n) \stackrel{\text{def}}{=} D(\delta f)$. By substitution of $\varepsilon$ with $\mu_a$, and using properties of variance, one may obtain the following expression for the variance of $\mu_a$:

$$D(\mu_a) = \frac{\sigma^2\cdot D(\delta f)}{A^2\cdot \sum_{i=1}^{n}z_i^2\cdot e^{-z_i^2}} \quad (8)$$

In general, (8) is sufficient to calculate error in determining $\mu_a$ when approximating data with a Gaussian function. Nevertheless, the expression can be made more convenient for calculations if the following conditions are met:

$$x_{i+1} - x_i = \Delta x \;\forall\; i: 0 < i < n \quad (9)$$

$$n \gg \frac{x_n - x_1}{\sigma} \quad (10)$$

$$x_1 < \mu \text{ and } \mu - x_1 \gg \sigma \quad (11)$$

$$x_n > \mu \text{ and } x_n - \mu \gg \sigma \quad (12)$$

Conditions (9) and (10) allow to turn the sum in (8) into the integral:

$$\sum_{i=1}^{n}z_i^2\cdot e^{-z_i^2} \xrightarrow{n\to+\infty} \frac{1}{\Delta x}\cdot\int_{x_1}^{x_n}\left(\frac{x-\mu}{\sigma}\right)^2 e^{-\left(\frac{x-\mu}{\sigma}\right)^2}dx \quad (13)$$

Conditions (11) and (12), in essence, mean that the measuring range covers the whole Gaussian peak. Integration from $x_1$ to $x_n$ can therefore be substituted with integration over the whole real line. The analytical solution to such an integral can easily be found and can be shown to be equal to $\frac{1}{\Delta x}\cdot\frac{\sqrt{\pi}}{2}\sigma$. Then, (8) is converted to an expression more convenient for calculations:

$$D(\mu_a) = \frac{2}{\sqrt{\pi}}\cdot\frac{D(\delta f)}{A^2}\cdot\sigma\cdot\Delta x \quad (14)$$

Expression (14) was actually used to calculate error in determining $\mu_a$ when approximating data by the least squares method in this work. For calculation of $D(\delta f)$, the author uses unbiased estimate of variance with unknown mean:

$$D(\delta f) = \frac{1}{n-1}\cdot\sum_{i=1}^{n}\left(\delta f_i - \overline{\delta f}\right)^2 \quad (15)$$

APPENDIX B

METHODOLOGY FOR DETERMINING TIME DELAYS BETWEEN LASER DIODES

Let $X$ and $Y$ be two arbitrary quantum states from set $\{P_\mu\}$, where $P$ denotes one of four possible polarization states and subscript $\mu$ denotes intensity at transmitter's source. Let $\delta t(Y, X)$ denote the difference in firing time between the laser diodes designed to generate states $X$ and $Y$. Let $t(X \to Z)$ and $t(Y \to Z)$ denote detection time, minus an integer number of periods, of states $X$ and $Y$ at an arbitrarily chosen detector $Z$. Then $\delta t(Y, X)$, which characterizes the degree of mismatch between the two lasers, can be written as:

$$\delta t(Y, X) = t(Y \to Z) - t(X \to Z). \quad (1)$$

For instance, the time delay between lasers $V_d$ and $V_s$ should be sought by analyzing the count statistics of detector $V$ since it obviously gets the most photons having vertical polarization, which ensures the best signal-to-noise ratio in this case:

$$\delta t(V_d, V_s) = t(V_d \to V) - t(V_s \to V). \quad (2)$$

In this work, the firing time of laser $H_s$ is assumed to be zero and the other laser diode delays are calculated relative to it. Thus, the time delay of $H_d$ relative to $H_s$ can be calculated as follows:

$$\delta t(H_d, H_s) = t(H_d \to H) - t(H_s \to H). \qquad (3)$$

It is somewhat more complicated to find the delay between $V_s$ and $H_s$ since these polarization states are orthogonal. If the difference in detection time of detectors $V$ and $H$ were zero, it could be calculated as:

$$\delta t(V_s, H_s) = t(V_s \to V) - t(H_s \to H) \qquad (4)$$

However, the response times of $V$ and $H$ are slightly different due to some design features of the ground receiver used in the experiments. Therefore, this discrepancy, which can be denoted as $\delta t(V, H)$, must also be taken into consideration:

$$\delta t(V_s, H_s) = t(V_s \to V) - t(H_s \to H) - \delta t(V, H) \qquad (5)$$

A time delay between two detectors can be found by comparison of their trigger times from the same quantum states. In this work, signal states at diagonal basis are used as such states to find the delay between $V$ and $H$:

$$\delta t(V, H) = t(D_s/A_s \to V) - t(D_s/A_s \to H) \qquad (6)$$

From (5) and (6), one can obtain:

$$\delta t(V_s, H_s) = t(V_s \to V) - t(H_s \to H) -$$
$$- t(D_s/A_s \to V) + t(D_s/A_s \to H) \qquad (7)$$

The delay between $V_d$ and $H_s$ can be calculated as:

$$\delta t(V_d, H_s) = \delta t(V_d, V_s) + \delta t(V_s, H_s) \qquad (8)$$

By substitution of the corresponding expressions from (2) and (7) into (8), one can get:

$$\delta t(V_d, H_s) = t(V_d \to V) - t(H_s \to H) -$$
$$- t(D_s/A_s \to V) + t(D_s/A_s \to H) \qquad (9)$$

The time delay between $D_d$ and $H_s$ can be found as:

$$\delta t(D_d, H_s) = t(D_d \to H) - t(H_s \to H) \qquad (10)$$

The delay between $A_d$ and $H_s$ can be found in a similar way:

$$\delta t(A_d, H_s) = t(A_d \to H) - t(H_s \to H) \qquad (11)$$

The time delay between $D_s$ and $H_s$ can be calculated as $\delta t(D_s, D_d) + \delta t(D_d, H_s)$. The first term can be found as $t(D_s \to D) - t(D_d \to D)$. If taking into account (10), one can thus obtain:

$$\delta t(D_s, H_s) = t(D_s \to D) - t(D_d \to D) +$$
$$+ t(D_d \to H) - t(H_s \to H) \qquad (12)$$

Similarly, the delay between $A_s$ and $H_s$ can be calculated as follows:

$$\delta t(A_s, H_s) = t(A_s \to A) - t(A_d \to A) +$$
$$+ t(A_d \to H) - t(H_s \to H) \qquad (13)$$

As one can notice, all the expressions for time delays contain at least one term like $t(P_s \to P)$. In accordance with the decoy-state BB84 protocol, Alice does not reveal what polarization she prepared for signal states: she only discloses the basis. Consequently, strictly speaking, Bob cannot get count statistics for $t(P_s \to P)$, he only possesses distribution for $t(P_s/OP_s \to P)$, where $OP_s$ is the orthogonal polarization state. However, the ground receiver has sufficiently high polarization exctinction ratios: the value averaged over all four polarization states does not fall below 150:1 during a whole communication session. Therefore, $t(P_s \to P)$ is equal to $t(P_s/OP_s \to P)$ with good accuracy. Accordingly, expressions (3,7,9–13) can be substituted with the following ones:

$$\delta t(H_d, H_s) = t(H_d \to H) - t(H_s/V_s \to H) \qquad (14)$$

$$\delta t(V_s, H_s) = t(H_s/V_s \to V) - t(H_s/V_s \to H) -$$
$$- t(D_s/A_s \to V) + t(D_s/A_s \to H) \qquad (15)$$

$$\delta t(V_d, H_s) = t(V_d \to V) - t(H_s/V_s \to H) -$$
$$- t(D_s/A_s \to V) + t(D_s/A_s \to H) \qquad (16)$$

$$\delta t(D_d, H_s) = t(D_d \to H) - t(H_s/V_s \to H) \qquad (17)$$

$$\delta t(A_d, H_s) = t(A_d \to H) - t(H_s/V_s \to H) \qquad (18)$$

$$\delta t(D_s, H_s) = t(D_s/A_s \to D) - t(D_d \to D) +$$
$$+ t(D_d \to H) - t(H_s/V_s \to H) \qquad (19)$$

$$\delta t(A_s, H_s) = t(D_s/A_s \to A) - t(A_d \to A) +$$
$$+ t(A_d \to H) - t(H_s/V_s \to H) \qquad (20)$$

Equations (14–20) are used to calculate laser diode time delays in this work.